\documentstyle[11pt,aaspp4,psfig,flushrt,tighten]{article}
\newcommand{\be}{\begin{equation}}  \newcommand{\ba}{\begin{eqnarray}}
\newcommand{\ee}{\end{equation}}  \newcommand{\ea}{\end{eqnarray}}
\newcommand{\etal}{et al.\ }  \def\gtsima{$\; \buildrel > \over \sim
\;$} \def\ltsima{$\; \buildrel < \over \sim \;$}
\def\gsim{\lower.5ex\hbox{\gtsima}}
\def\lsim{\lower.5ex\hbox{\ltsima}}
\def\simgt{\lower.5ex\hbox{\gtsima}}
\def\simlt{\lower.5ex\hbox{\ltsima}}
\def\simpr{\lower.5ex\hbox{\prosima}}

  \def\msun{{M_\odot}}
  \def\eg{{\frenchspacing\it e.g. }}

\begin{document}
\title{Toward an Improved Analytical Description of  Lagrangian Bias}

\author{Evan Scannapieco\altaffilmark{1} \& Robert J.
Thacker\altaffilmark{2}} \altaffiltext{1}{ Kavli Institute for
Theoretical Physics, Kohn Hall, UC Santa Barbara, Santa Barbara, CA
93106} \altaffiltext{2}{Department of Physics and Astronomy, McMaster
University, 1280 Main St.\ West, Hamilton, Ontario, L8S 4M1, Canada.}

\begin{abstract}

We carry out a detailed numerical investigation of the spatial
correlation function of the initial positions of cosmological dark
matter halos.  In this Lagrangian coordinate system, which is
especially useful for analytic studies of cosmological feedback, we
are able to construct cross-correlation functions of objects with
varying masses and formation redshifts and compare them with a variety
of analytical approaches.   For the case in which both formation
redshifts are equal,  we find good agreement between our numerical
results and the bivariate model of  Scannapieco \& Barkana (2002;
SB02) at all masses, redshifts, and separations, while the model of
Porciani \etal (1998) does well for all parameters except for objects
with different masses at small separations.  We find that the standard
mapping between Lagrangian and Eulerian bias performs well for rare
objects at all separations, but fails if the objects are
highly-nonlinear (low-sigma) peaks.  In the Lagrangian case in which
the formation redshifts differ, the SB02 model does well for all
separations and combinations of masses, apart from a discrepancy at
small separations in situations in which the smaller object is formed
earlier and the difference between redshifts or masses is large.  As
this same limitation arises in the standard approach to the
single-point progenitor distribution developed by Lacey \& Cole
(1993), we conclude that a more complete understanding of the
progenitor distribution is the most important outstanding issue in the
analytic modeling of Lagrangian bias.

\end{abstract}

\keywords{cosmology: theory --  large-scale structure of the universe}

\newpage 
\section{Introduction}

The universe is a complicated place.  Not only does cosmological
structure formation involve detailed questions of fluid dynamics,
radiative transfer, and chemistry, but most of these issues are
non-local.  The impact of high-redshift supernovae (SNe), for example,
is not limited to the host-galaxies in which they are
born. Theoretical arguments show that the large numbers of SNe formed
in high redshift dwarf galaxies drove global winds of such force that
they not only reduced further infall onto their hosts, but also
suppressed the formation of neighboring galaxies (\eg Scannapieco,
Ferrara, Broadhurst 2000; Thacker, Scannapieco, \& Davis
2003). Observational support of this idea is strong, with large
numbers of massive outflowing starbursts directly detected at $z
\gtrsim 3$ (Pettini \etal 2001; Frye, Broadhurst, \& Benitez 2002; Hu
\etal 2002).

Such  observations of outflows are complemented by hints that the
intergalactic medium (IGM) surrounding high-redshift starbursts has
been excavated by outflows out to distances $\sim 1$ Mpc (Adelberger
\etal 2003) and that the distribution of intergalactic metals is
biased and highly clustered around similar sources (Pichon \etal
2003).  Furthermore, quasar absorption line studies have uncovered an
inhomogeneous distribution of heavy elements that is already pervasive
throughout the intergalactic medium by $z \approx 5$ (\eg Rauch,
Haehnelt, \& Steinmetz 1997; Songaila 2001; Schaye \etal 2003; Aracil
\etal 2003).  This large-scale enrichment further complicates the
structure formation  process.  Metal-line cooling is of crucial
importance to the formation history of large galaxies (\eg Sutherland
\& Dopita 1993), as low levels of enrichment can produce an order of
magnitude change in cooling rates above $10^5$ K. Also, the presence
of even trace levels of heavy elements in protogalactic material may
be sufficient to drastically affect the shape of the stellar initial
mass function (\eg Bromm \etal 2001; Schneider \etal 2002).

But the most highly non-local processes are undoubtedly radiative.
The earliest such example is the dissociation of primordial H$_2,$
which regulates the formation of the first generation of galaxies in
halos with virial temperatures below $10^4$ K.  As ${\rm H}_2$ is
easily photo-dissociated by 11.2-13.6 eV photons to which the universe
is otherwise transparent, even small numbers of forming stars have the
potential to suppress dwarf galaxy formation in large regions of space
(\eg Haiman, Rees, \& Loeb 1997), motivating detailed analysis of
${\rm H}_2$ formation by X-rays (\eg Oh 2001) and self-shielding
(Yoshida \etal 2003).  Secondly, there is the question of
reionization.  Here the classic picture is of a two-stage process,
which begins with sources ionizing their immediate surroundings and
ends in a rapid ``overlap'' phase as H~II regions join together,
quickly ionizing the remaining neutral regions (\eg Gnedin 2000).  At
these times feedback due to photo-evaporation is of primary importance
to galaxy formation in $T_{\rm vir} = 10^4-10^{4.5}$K halos (\eg
Barkana \& Loeb 1999), a non-local effect that is only worsened by the
fact that similar low-mass galaxies are likely to be the source of the
ionizing background in the first place.

This wide range of mechanical, chemical, and radiative feedback
processes has necessarily sparked a multifaceted numerical attack on
non-local cosmological problems. Recent numerical simulations have
included detailed studies of mechanical feedback from galaxy outflows
(\eg Scannapieco, Thacker, \& Davis 2001; Springel \& Hernquist 2003),
the chemical properties of galaxies and the intergalactic medium (\eg
Thacker, Scannapieco, \& Davis 2002; Theuns \etal 2002; Tassis \etal
2002) and the formation and propagation of cosmological ${\rm H}_2$
dissociation and ionization fronts (\eg Ricotti, Gnedin, \& Shull
2002; Ciardi, Ferrara, \& White 2003; Yoshida \etal 2004).  These
highly non-local studies, unlike earlier investigations, are far
outstripping the power of approximate analytic techniques.  Thus, in
our rush to match the dramatic recent observational advances in
cosmology with equally sophisticated numerical work, there is a very
real threat that we will soon lack the tools to understand  either in
a more fundamental context.

The most widely applied analytic method for determining the
distribution of halos was first developed by Press \& Schechter (1974,
hereafter PS), and later refined by Bond \etal (1991), Lacey \& Cole
(1993), and Sheth, Mo, \& Tormen (2001, hereafter SMT01) among others.
Yet, despite its widespread utility, this method is limited in that it
can only predict the average number density of virialized dark matter
halos, and does not supply any information about their relative
positions. However, when used in combination with simulations, this
model is exceptionally useful since it allows rapid investigation of
parameter space. In the absence of this possibility for studying
non-local effects, numerical investigations are thus on a surprisingly
shaky footing: any such investigation is computationally intensive,
and is able to consider only a limited range of possible scenarios and
parameters. Yet, at the same time, most processes depend on a large
number of uncertain parameters, almost assuring that the simulations
will not be able to bracket the full range of physically plausible
scenarios.  Finally, even in the cases in which simulations are able
to assess these uncertainties, the lack of an analytical counterpart
frequently leaves us  without a basic theoretical understanding.

These limitations are somewhat mitigated by analytical studies of the
two-point correlation function,  although only recently  have these
approaches been  extended to the point of being useful for  feedback.
Considerable effort has gone into to modeling the biased clustering of
halos at a given epoch in the Eulerian coordinate system in which they
are observed, both in mildly-nonlinear (see Bernardeau \etal 2002 for a
review) and highly nonlinear contexts (\eg Mo \& White 1996, hereafter
MW96; Seljack 2000; Ma \& Fry 2000). Limited models have also been
developed in the Lagrangian coordinate system in which the halos
cell-centers were originally located (Porciani \etal 1998; Sheth \&
Lemson 1999).  In this  reference frame, the distance between two
objects is closely related to the total mass of material between them,
rather than their final comoving distance.

The usefulness of such a reference frame is twofold.  Firstly, it is the
natural frame for PS-type analytical calculations, which  associate
peaks  in the {\em initial} density field with collapsed objects at
various redshifts.  Secondly, it allows for a more accurate treatment
of the propagation of disturbances such as shocks (Scannapieco \etal
2003) or ionization fronts (Iliev \etal 2004) between objects, as
these are largely dependent on the total column depth of material
separating two perturbations  rather than their precise distance in
physical space.   However, early analytical approaches were unable
capture the Lagrangian clustering between halos forming at different
epochs, a quantity of crucial importance to modeling non-local
effects, as cosmological disturbances take a significant amount of
time to propagate from their sources to neighboring objects.

In an attempt to alleviate this shortcoming, we developed in
Scannapieco \& Barkana (2002; hereafter SB02) an approximate
analytical model that considered the collapse of two neighboring
points of arbitrary mass and formation redshift, separated by an
arbitrary Lagrangian distance. Working in the context of the linear
excursion-set formalism of Bond \etal (1991), we showed that while the
exact two-point solution could not be obtained analytically, this
solution could be reproduced to high accuracy by introducing a simple
approximation.  The resulting expressions then interpolated smoothly
between all standard analytical limits: reducing for example to the
standard halo bias expression described in MW96 in the limit of
equal-mass halos at the same redshift, and reproducing the Lacey \&
Cole (1993) progenitor distribution in the limit of different-mass
halos at the same position and different redshifts.

However, just as numerical results must be placed in a larger
analytical context, the analytical results obtained in SB02 were  left
incomplete without an accurate numerical assessment of their strengths
and limitations.  In this study, we carry out just such a comparison,
computing detailed numerical Lagrangian halo-halo correlations at a
variety of masses and redshifts and comparing them directly with SB02
and other analytical models.  The structure of this work is as
follows. In \S 2 we describe our numerical simulation and group
finding techniques and compare the resulting mass distributions with
standard analytical results.  In \S 3 we systematically compare the
Lagrangian correlation function of groups with the analytical model of
SB02 for a  variety of formation redshifts and halo masses. We
conclude in \S 4 with a  brief review of our results and a short
discussion.

\section{Simulations and Group Finding}

To achieve high accuracy at small separations while at the same time
minimizing the effects of box-mode damping, we conducted two detailed
numerical simulations using a parallel OpenMP-based version of the
HYDRA code (Couchman \etal 1995; Thacker \& Couchman 2000).   Our
assumed cosmological parameters were taken to correspond to the
generally accepted ``concordance'' values, which agree well with
measurements of the Cosmic Microwave Background, the number abundance
of galaxy clusters, and high-redshift supernova distance estimates
(\eg Spergel \etal 2003; Eke \etal 1996; Perlmutter \etal 1999).  In
this case $h=0.65$, $\Omega_0$ = 0.3, $\Omega_\Lambda$ = 0.7,
$\Omega_b = 0.05$, $\sigma_8 = 0.87$, and $n=1$, where $\Omega_0$,
$\Omega_\Lambda$, and $\Omega_b$ are the total matter, vacuum, and
baryonic densities in units of the critical density, $\sigma_8^2$ is
the variance of linear fluctuations on the $8 h^{-1}{\rm Mpc}$ scale,
and $n$ is the ``tilt'' of the primordial power spectrum.  We used the
Bardeen \etal (1986) transfer function with an effective shape
parameter of $\Gamma=0.18$.

The first run (Run A), chosen to do best at small distances, was
carried out in a cubic volume $78.5$ comoving Mpc on a side, populated
with $350^3$ dark matter particles. This run is a continuation of that
used in Scannapieco \& Thacker (2003). The second run (Run B), chosen
to do best at large separations, was carried out in a cubic volume
$113$ comoving Mpc on a side, populated with $400^3$ dark matter
particles.  The mass of each particle was $3.9 \times 10^{8} \msun$
for Run A and $7.9 \times 10^{8} \msun$ for Run B. Since we restrict
our results to masses above $10^{11} \msun$ we have over 250 and 125
particles in each group used in the study.  Both runs were integrated
from an initial redshift of $z=49$ down to $z=1$ and used fixed
physical Plummer softening lengths of 5.7 kpc (Run A) and 6.9 kpc (Run
B).  Both simulations were performed with 64-bit precision.

Group identification was carried out using the HOP (Eisenstein \& Hut
1998) algorithm, which uses the local density for each particle to
trace  (``hop'') along a path of increasing density to the nearest
local maximum,  at  which point the particle is assigned to the group
defined by this maximum.  Since this process assigns all particles to
groups, many of these must be removed by requiring an outer threshold
density, $\delta_{\rm outer}$.  Finally, a ``regrouping'' stage is
needed in which all groups are merged for which the boundary density
between them exceeds $\delta_{\rm saddle}$, and all objects thus
identified must have one particle whose density exceeds $\delta_{\rm
  peak}$ to be accepted as a final group (see Eisenstein \& Hut 1998
for explicit details).

\begin{figure}
\centerline{\psfig{file=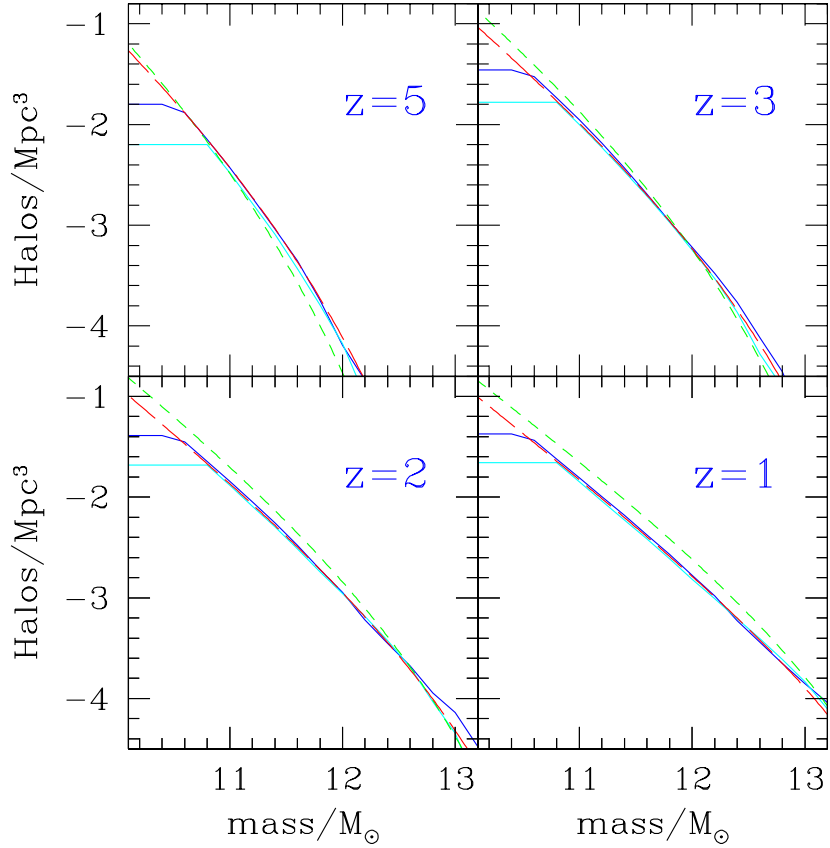,width=5.0in}}
\caption{Comparison of numerical and analytic mass functions.   Here
the solid lines extending to lower masses  are calculated from the
simulation Run A, the solid lines that have a higher low-mass limit
are calculated from simulation Run B, and the short-dashed and
long-dashed lines represent the PS and the SMT01  models,
respectively.  Good agreement between the results of both simulations
and SMT01 is found at all redshifts.}
\label{fig:mass}
\end{figure}

Applying these criteria with the HOP parameters  $N_{\rm dens}=48$,
$N_{\rm hop}=20$, $N_{\rm merge}=5$, $\delta_{\rm peak}=160$,
$\delta_{\rm saddle}=140$, and $\delta_{\rm outer}=80$, yielded  1548,
4841, 6179 and 6739 groups with masses above $10^{11} \msun$ at
redshifts $z=5$, 3, 2, and 1, respectively in Run A, and 4276, 13118,
17063,and 18424 at the same redshifts in Run B.  The resulting
cumulative mass functions for both runs are shown in Figure
\ref{fig:mass}, where they are compared with the standard analytical
PS mass function and the refined fit from the ellipsoidal collapse
model of SMT01.  While the PS model exhibits substantial differences
from the simulation group densities, particularly for the
late-forming, low-mass peaks, the SMT01 expression provides a good fit
for both simulations over the full range of masses and redshifts of
consideration, as has been seen in several previous comparisons (Bode
2001; Jenkins \etal 2001).  Thus we expect our groups to accurately
represent a ``standard'' sample as defined in the majority of  recent
N-body simulations.

To define the Lagrangian coordinates of a group, we trace back the
position of each of the particles contained within it to the start of
the numerical simulation, and then compute the center of mass of this
distribution.   Using these positions we are then able to numerically
construct $\frac{d^2 n}{dM_1 dM_2} (M_1,z_1, M_2, z_2, r)$ as
described in SB02 eqs.\ (41) and (42),   the distribution function of
halos at any two given masses ($M_1$ and $M_2$), output redshift
($z_1$ and $z_2$), and {\em initial} comoving separation $r$.  In
practice however, it is more convenient to normalize this function by
the average number of halos at each mass and redshift, defining the
Lagrangian correlation function $\xi_L$ as the excess probability of
finding two halos that are initially separated by a comoving distance
$r$.  That is \be \xi_L(M_1,z_1, M_2, z_2,r) + 1 \equiv
\left[\frac{d^2 n}{dM_1 dM_2} (M_1,z_1, M_2, z_2, r) \right]
\left[\frac{dn}{dM_1} (M_1,z_1) \frac{dn}{d M_2} (M_2,z_2)
\right]^{-1}, \ee where $\frac{dn}{dM}$ is the overall distribution of
halos at a single  mass and redshift.  It is this excess, then, that
we study in detail in  the comparisons presented below.

\section{Comparisons with Analytics}

\subsection{Single Redshift, Single Mass}

In order to study the behavior of the Lagrangian correlation function
over a range of masses, we divided our sample of groups into  3 bins
at each redshift, containing objects of total mass  $10^{11} \msun
\leq M < 10^{11.5} \msun$,  $10^{11.5} \msun \leq M < 10^{12} \msun$,
and $10^{12} \msun \leq M < 10^{12.5} \msun$. While the central
values in each of these bins are  $10^{11.25} \msun$, $10^{11.75}
\msun$, and $10^{12.25} \msun$ respectively, we see from the steep
mass functions in Figure  \ref{fig:mass} that the majority of groups
in each bin are dominated by the smallest values.  In fact, for all
bins and redshift ranges considered here, the mean is well
approximated by the minimum value plus a third of the width.  Thus,
throughout this study, we refer to these subsets as bins of mass
$10^{11.15} \msun$,  $10^{11.65} \msun$, and $10^{12.15} \msun$,
comparing them with analytic results for these values.

\begin{figure}
\centerline{ \psfig{file=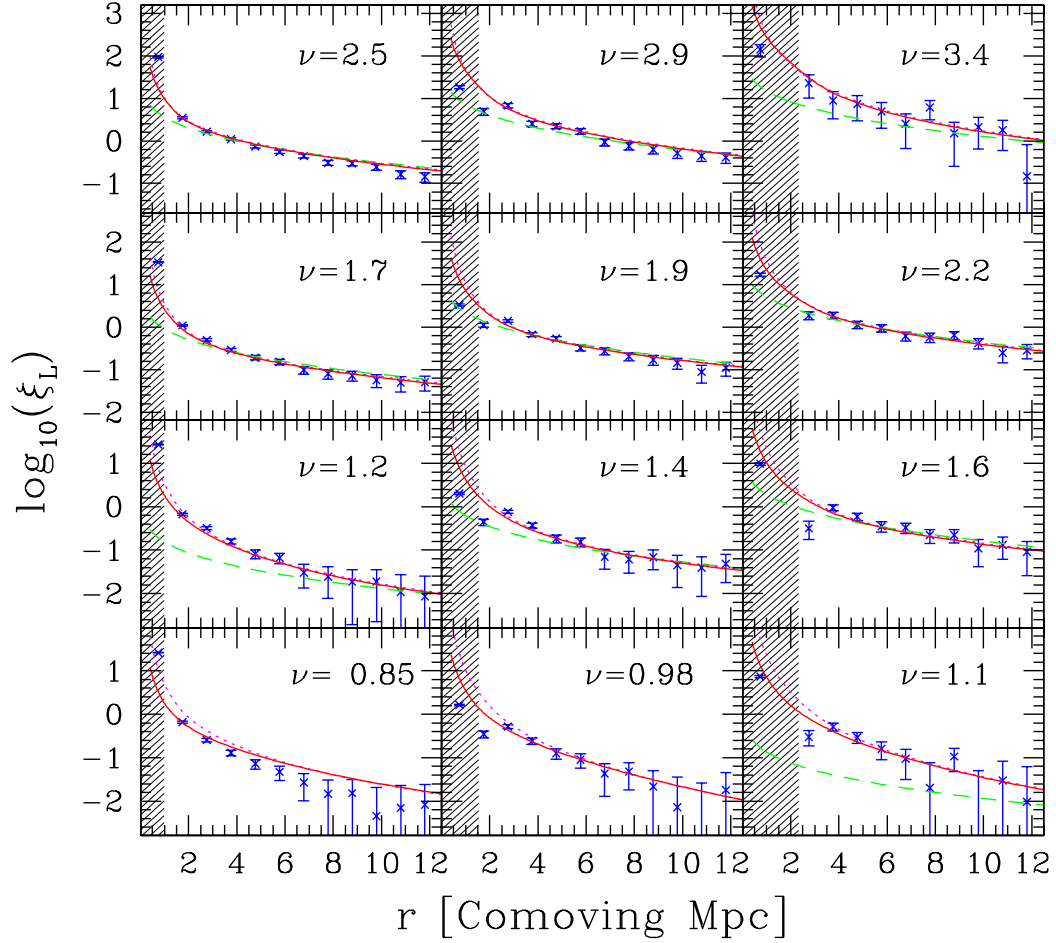,width=6.2in}}
\caption{Lagrangian correlation function of objects with the same mass
at the same redshift.  From top to bottom $z = 5$, $z = 3$, $z = 2$,
and $z = 1$, and from left to right the mass bins are centered on
$10^{11.15} \msun,$ $10^{11.65} \msun$ and $10^{12.15} \msun$.   The
points are the simulation values, the solid lines are the SB02 model,
the dotted lines are the Porciani \etal (1998) model and the dashed
lines are the simple MW96  geometric bias estimates.  
The shaded regions are within the (spherical) radius of the halo 
as defined by eq.\ \ref{eq:RL} .We omit the MW96
estimate in cases in which $\nu \leq 1$.}
\label{fig:xiall}
\end{figure}

Constructing $\xi_L$ for each of these mass bins and redshifts
resulted in the points depicted in Figure \ref{fig:xiall}, in which we
have taken a bin width of 1 Mpc.  We have computed error bars
including both the standard (Poissonian) error, which depends on the
number of pairs in a given bin, as well as the additional scatter
caused by the finite sample size used to construct the correlation
function.  In this case the variance in the number of pairs in a given
bin $i$ in a single simulation is well-approximated by \be
\sigma^2(N_{i,1,2})  = N_{i,1,2} + N_{i,1,2}^2
\left[\frac{2}{N_1}+\frac{2}{N_2} \right], \ee where $N_{i,1,2}$ is
the number of pairs in a bin $i$, while $N_1$ and $N_2$ are the total
numbers of objects of  each of the two types being measured (Mo, Jing,
\& B\" orner 1992).  Thus combining two measurements of $\xi_L$ with
relative weights $w^A$ and $w^B$ leads to a total error in each bin
$i$ and objects types $1$ and $2$ of  \be \sigma^2(\xi^{\rm
Tot}_{L,i,1,2}) = (w^A_{1,2})^2 (\xi^A_{L,i,1,2}+1)^2
\sigma^2(N^A_{i,1,2}) + (w^B_{1,2})^2 (\xi^B_{L,i,1,2}+1)^2
\sigma^2(N^B_{i,1,2}), \ee where we choose our weights based on the
total number of pairs with masses 1 and 2 in each simulation at any
given output $z$: \ba w^A_{1,2} &=& N^A_{1} N^A_{2}/(N^A_{1} N^A_{2}+
N^B_{1} N^B_{2}), \nonumber \\ w^B_{1,2} &=& N^B_{1} N^B_{2}/(N^A_{1}
N^A_{2}+ N^B_{1} N^B_{2}).  \ea

Each panel in Figure \ref{fig:xiall} is labeled by its ``density
threshold,'' $\nu$, a value that arises in the analytical mass
functions.  This is defined as \be \nu \equiv 1.69 D(z)^{-1}
\sigma(M)^{-1}, \ee where $D(z)$ is the linear growth factor.  The
density threshold can  then be used to construct a standard
``geometrical bias'' estimate of the correlation function: \be
\xi_L(M_1,z_1,M_2,z_2,r)  \approx b_L(M_1,z_1) b_L(M_2,z_2) D(z_1)
D(z_2) \xi_{\rm DM}(r), \ee where $b_L \equiv (\nu^2-1)/1.69$ and
$\xi_{\rm DM}$ is the underlying matter correlation function, linearly
extrapolated to the present epoch (\eg Kaiser 1984; MW96).  For
reference this estimate of $\xi_L$ is shown as the short-dashed lines
in each panel, while the solid lines give the more detailed model
developed in SB02.  Finally, the dotted lines are from the model
described in eq.\ (A1) of Porciani \etal (1998; hereafter P98), which
is similar to SB02, but applicable only in the case of two halos with
the same collapse redshift.

Surveying this figure, we see that, in general, there is good
agreement  between the SB02 model and the numerical results at all
masses and redshifts.  Furthermore, in the case of $M_1 = M_2$ the P98
model is nearly equivalent to this expression, and thus obtains
similar agreement.  On the other hand, the standard MW96 model, which
was derived in the limited case of large separations and high values
of $\nu$, is not able to trace $\xi_L$ within a distance of $\sim 4$
comoving Mpc, or $\nu$ values less than 1.5.  Note that for $\nu < 1$
the MW96 model becomes negative, and thus we omit this simple estimate
for these cases.

Even in the more detailed SB02 model, a minor discrepancy is seen,
namely a deficit in the numerical results in the 1-2 Mpc bins. The
difference is most  noticeable in the $10^{12.15} \msun$ case, where
the 2 Mpc bin is missing  from all plots. The explanation of this
phenomenon is   straightforward and arises from the group finding
process.  For our chosen cosmology,  a (spherical) Lagrangian region
that encompasses a mass $M$, will have a  comoving radius, $R_L(M)$,
that is given by   \be  R_L(M) = 1.9 \, \left[ M/(10^{12} \msun)
\right]^{1/3}  \qquad {\rm Mpc}.
\label{eq:RL}
\ee Thus in the $10^{12.15} \msun$ plot the entire second bin is
contained within  the value of $R_L$, while for the $10^{11.65} \msun$
case approximately  half of  the bin is contained within $R_L$.  It is
clear that the deficit is caused by the fact that HOP, like any
group-finding  algorithm, will consider objects at these separations
as a single higher-mass group, excluding them from the numerical
calculation of $\xi_L$.  Note that the correlation function at $0$
separation is formally infinite, as an object of mass $M$ at a
redshift $z$ is {\em always} found at a  distance of $r=0$ from an
object of similar mass and redshift (namely itself).  This means that
the innermost bin is also not interesting for our purposes.

\begin{figure}
\centerline{ \psfig{file=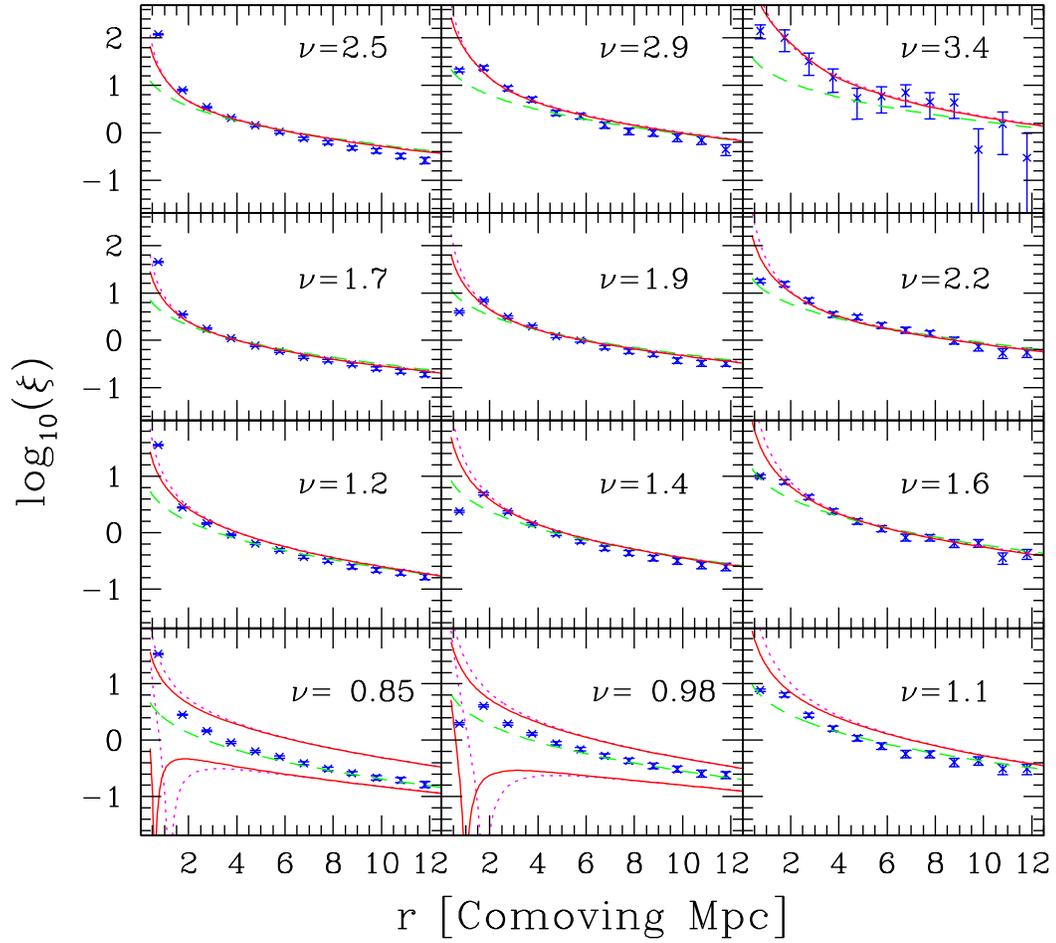,width=6.2in}}
\caption{Eulerian correlation function of objects with the same mass
at the same redshift.  Panels are as in Figure
\protect\ref{fig:xiall}.  The points are the  simulation values, and
the dashed lines are the simple MW96 Eulerian geometric bias
estimates.  The solid and dotted lines are the SB02 and P98 models as
mapped to Eulerian coordinates as described in the text. In the cases
in which $\nu \leq 1,$ both possible values for $b_E$  as in eq.\ (8)
are shown.}
\label{fig:xialleul}
\end{figure}

To compare our simulations with more standard estimates we plot the
Eulerian correlation function in comoving coordinates, for objects
with the same mass and redshift,  in Figure \ref{fig:xialleul}.  In
this case the bias factor that appears in the MW96 estimate is
modified to $b_E = (\nu^2-1)/1.69 +1 = b_L + 1$ to account for the
motion of the halos toward each other.  We find good agreement between
the estimates and numerical values at large separations, as has been
previously studied in detail (\eg Jing 1998), again giving us
confidence in our numerical sample.

In a numerical study of several $256^3$ simulations, Jing (1999) found
that the $b_E = b_L + 1$ mapping between Eulerian and Lagrangian
coordinates provided a good fit to these correlation functions at
large distances.  Furthermore, he conjectured that the nonlinear
effect in the mapping between coordinate systems was not a major
source of the inaccuracy of the MW96 formula.  To test this conjecture
in the context of our analytical modeling, we have also plotted in
Figure \ref{fig:xialleul} the bias estimated as \be b_E(z,r)  = 1 \pm
\sqrt{{\rm abs}[\xi_L(r) D(z)^{-2} \xi_{\rm DM}(r)^{-1}]},
\label{eq:bE}
\ee where  $\xi_L(r)$ is the Lagrangian correlation function as
calculated by SB02 or P98, and the $\pm$ sign is taken to be positive
if $\nu \geq 1$ and negative if $\nu < 1$.  Disregarding the innermost
bin, we find that for the $z=5$, 3, and 2 results, this conjecture
indeed seems to hold.  As $\nu$ approaches, and then falls below 1
however, this approximation breaks down.  Thus a more sophisticated
Eulerian to Lagrangian mapping is likely to be necessary to reproduce
the behavior of these correlation functions at small separations (as
in \eg Iliev \etal 2003).

\subsection{Single Redshift, Two Masses}

\begin{figure}
\centerline{ \psfig{file=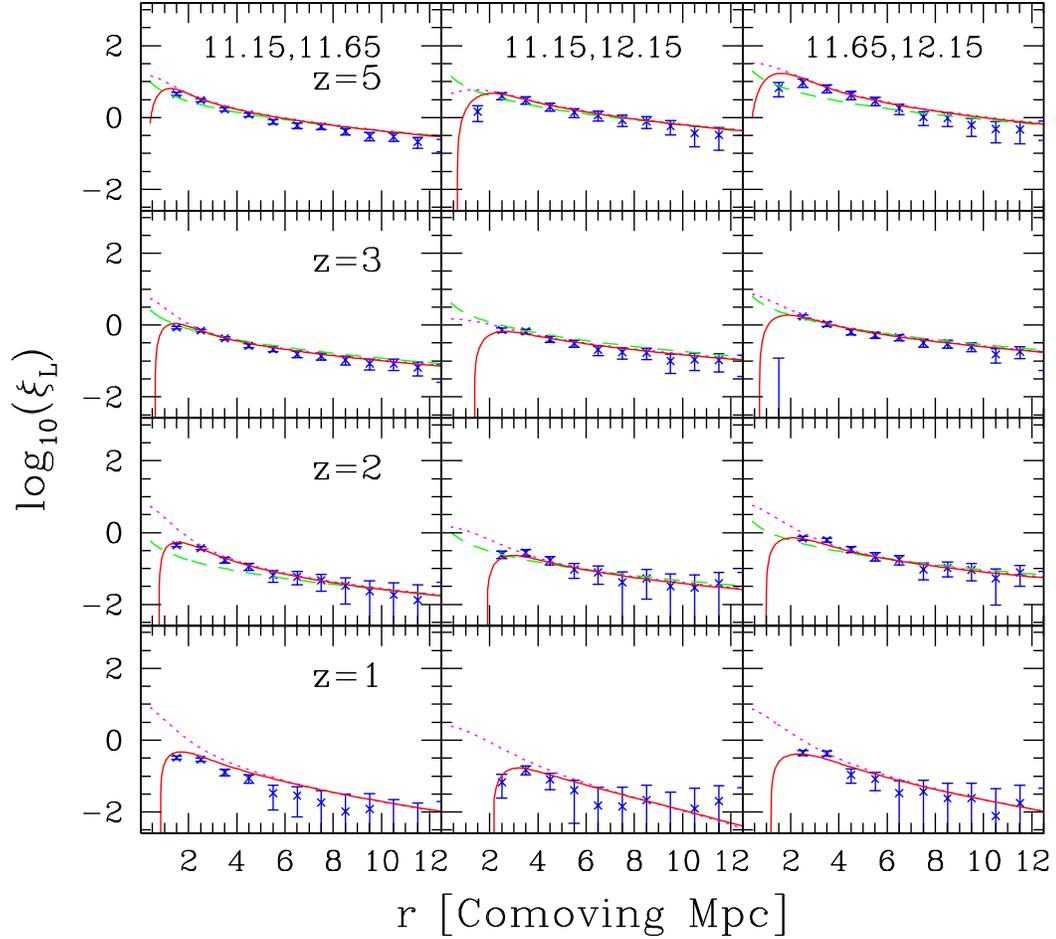,width=6.2in}}
\caption{Lagrangian correlation function of objects with different
masses at the same redshift.  From top to bottom $z = 5$, $z = 3$, $z
= 2$, and $z = 1$, and from left to right the mass bin pairs are
centered on ($10^{11.15} \msun$, $10^{11.65} \msun$), ($10^{11.15}
\msun$, $10^{12.15} \msun$), and ($10^{11.65} \msun$, $10^{12.15}
\msun$), respectively.  Lines and points are as in Figure
\protect\ref{fig:xiall}, with the MW96 estimate omitted if $\nu \leq
1$.}
\label{fig:xiall2}
\end{figure}

Next we turn our attention to the case of two dark matter halos with
different masses. In Figure \ref{fig:xiall2} we plot the correlation
function for mass bin pairs ($10^{11.15} \msun$,$10^{11.65}\msun$),
($10^{11.15} \msun$,$10^{12.15}\msun$) and ($10^{11.65}
\msun$,$10^{12.15}\msun$) at our chosen redshifts. As in  the
single-mass case, all analytical approaches do well at large
separations and high $\nu$ values.  Also as in the single-mass case,
the MW96 result fails when $\nu(M,z)$ approaches 1 for either of the
halos under comparison, while the P98 and SB02 results are able to
handle this range of parameters.  Unlike the single mass case,
however, only the SB02 model is able to reproduce the behavior at
small separations.  This is because at these distances the MW96 model
is too simplified, while the P98 model does not correctly implement
the ``barriers'' that exclude the formation of a density peak within a
larger peak in the excursion set formalism (Bond \etal 1991).  Indeed
in Figure 3 of P98, the authors point out that their approximation
fails to reproduce the exact excursion-set solution at separations $r
\lesssim 5 R_L(M),$ with $R_L$ defined as in eq.\ (\ref{eq:RL}).
These barriers are properly accounted for, however, in the SB02 model,
and thus the presence of two objects of different masses at the same
position and redshift is excluded in this formalism.

\begin{figure}
\centerline{ \psfig{file=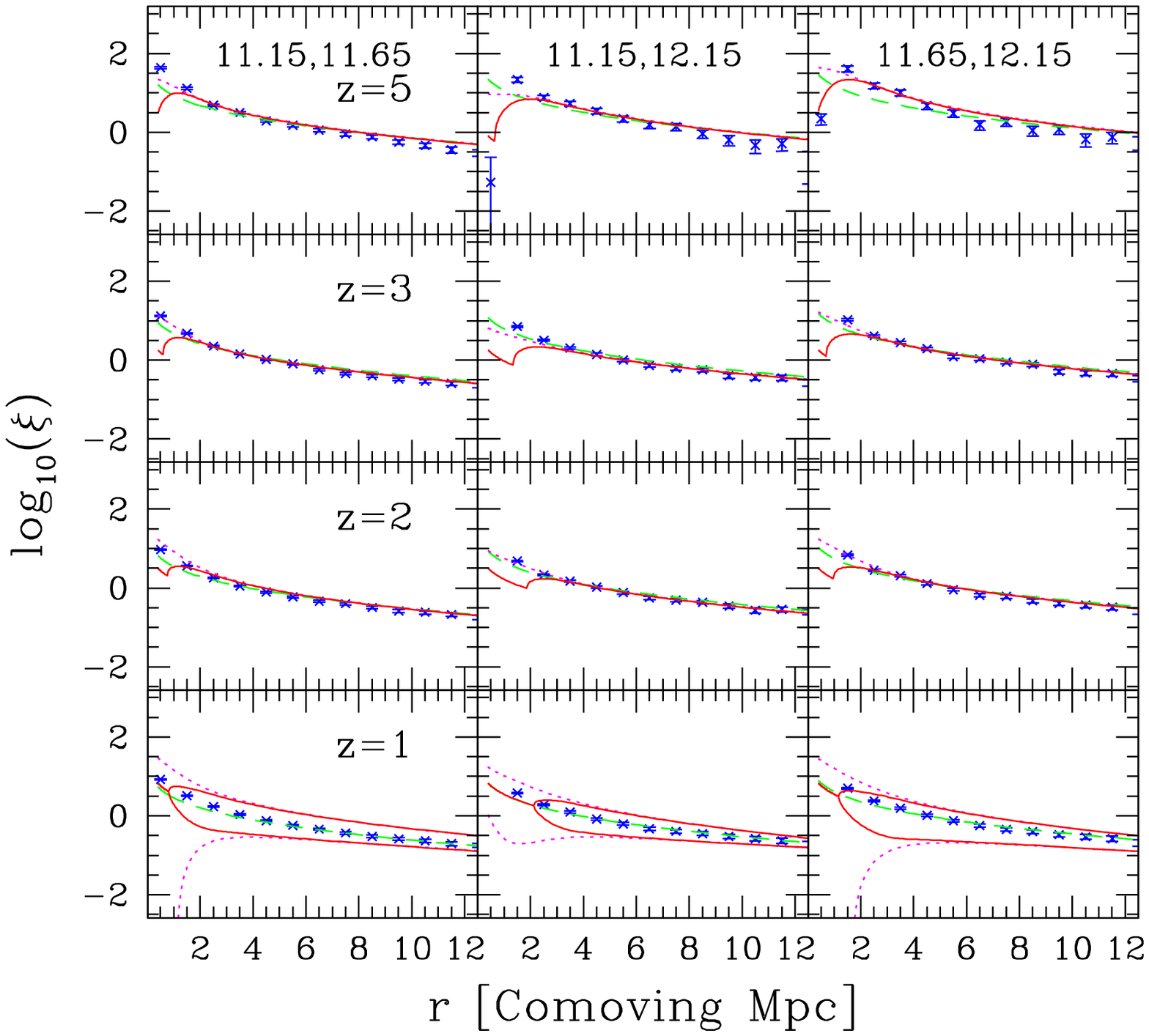,width=6.2in}}
\caption{Eulerian correlation function of objects with different
masses at the same redshift.  Panels are as in Figure
\protect\ref{fig:xiall2}.  The points are the  simulation values, and
the dashed lines are the simple MW96 Eulerian geometric bias
estimates.  The solid and dotted lines are the SB02 and P98 models as
mapped to Eulerian coordinates as described in the text. In the $z =
1$ cases both possible values for $b_E$  as in eq.\ (8) are shown.}
\label{fig:xialleul2}
\end{figure}

In Figure \ref{fig:xialleul2} we compare the Eulerian correlation
functions measured in our simulations with the ($b_E =
\nu^2-1/1.69+1$) MW96 estimates and the P98 and SB02 estimates as
mapped to this coordinate system by eq.\ (\ref{eq:bE}).  As in Figure
\ref{fig:xiall2}, all models do well at large separations, when $\nu >
1$.  Unlike the Lagrangian case, however, no exclusion is seen in the
numerical results at small separations.  As collapse and virialization
decrease the radius of a perturbation by a factor $\sim 1/6$, there is
nothing preventing two objects whose {\em initial} centers of mass
were quite separated from moving together to very close distances.
Thus the Eulerian version of the SB02 model is a poor approximation at
these separations, while the P98 and MW96 models, which do not impose
this exclusion, do as well in most cases.  In fact, the MW96 model,
despite its simplicity, provides the best overall fit to the data,
while both of the more sophisticated models fail at $z=1.$ While this
agreement provides us a simple approximation in Eulerian space, it
again serves to point out that the relationship between Eulerian and
Lagrangian spaces is complex for low $\nu$ values and small distances.

\subsection{Two Redshifts, Single Mass}

While the single redshift comparisons above give us good confidence in
our analytical modeling of simultaneously collapsing halos, this case
is only of secondary importance for non-local effects in structure
formation.  In reality, cosmological disturbances such as ionization
fronts and outflows take an appreciable amount of time to propagate
from their sources to neighboring objects.  In fact, the majority of
feedback effects are most efficient if they reach the neighboring
perturbations {\em before} they have virialized (\eg Haiman, Rees, \&
Loeb 1997; Scannapieco, Ferrara, Broadhurst 2000).  Thus it is
essential that any analytical treatment of non-local effects in
structure formation be capable of adequately handling these cases.

Two-redshift correlation functions are not unknown observationally.
The positions of galaxies measured at a redshift $z_1$ can in
principle  be correlated with the positions of galaxies at an earlier
redshift  $z_2.$    However, for both objects to be observed
simultaneously  and not be behind each other, $r > c [t(z_1)-t(z_2)]
,$ meaning that the light from the $z_2$ galaxy (as we see it today)
is unable to reach the $z_1$ galaxy (as we observe it).  In fact, such
distances are long indeed, $\gtrsim 1000$ comoving Mpc for  our output
times, and well beyond our ability to simulate.  Instead we
concentrate here on distances $r \ll c [t(z_1)-t(z_2)]$  which, while
not directly observable, are much more typical for the propagation of
cosmological disturbances.

\begin{figure}
\centerline{ \psfig{file=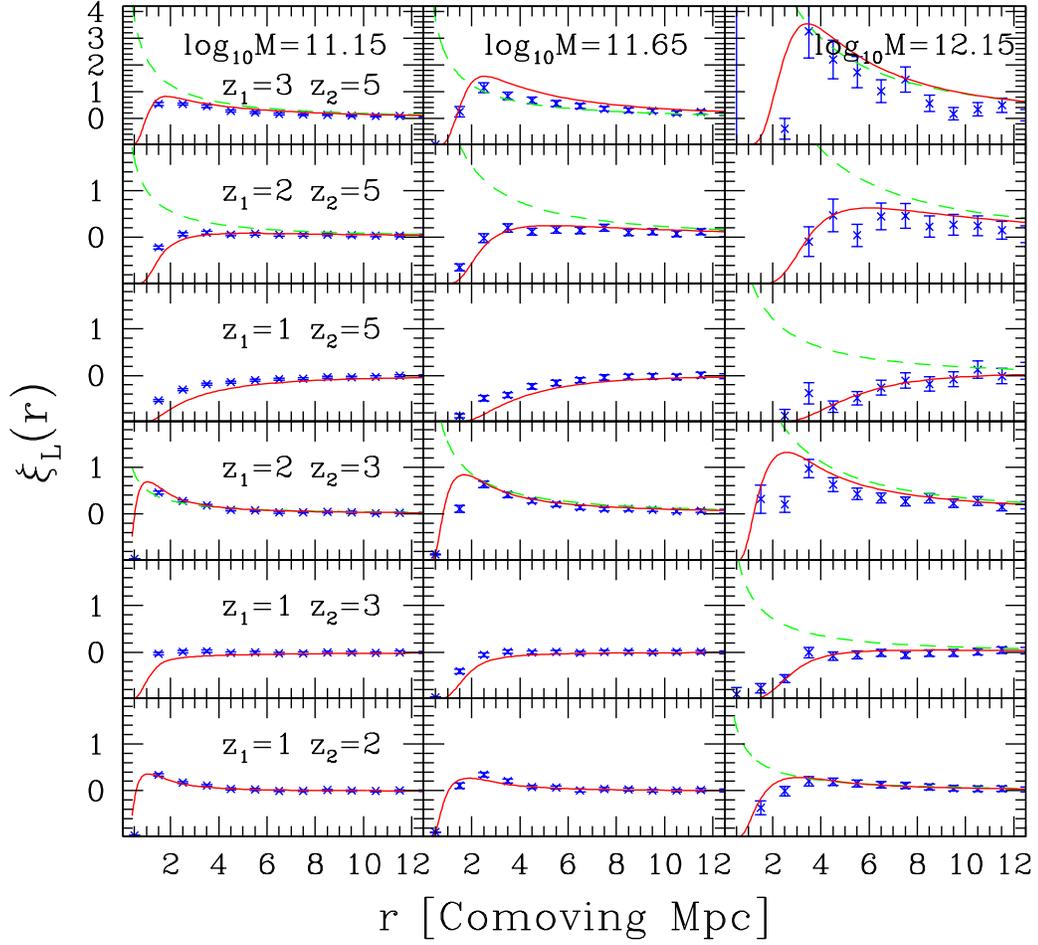,width=6.2in}}
\caption{Lagrangian correlation function of objects with the same
mass, at different redshifts. Each row is labeled by its $z$ values
and each column is labeled by its $M$ value.  In all panels the solid
lines are the SB02 model, and the dashed lines are the MW96 model in
the cases in which $\nu \geq 1$ for both halos.  Note that we plot
$\xi_L$ rather than $\log_{10}(\xi_L)$ as the correlation function
becomes negative at small separations.}
\label{fig:xiall3}
\end{figure}

In Figure \ref{fig:xiall3}, we again consider two halos with the same
mass, but now with different collapse redshifts.  At large
separations, the numerical values display much the same behavior as in
Figure \ref{fig:xiall}. The correlation between halos is stronger for
rarer objects, and weaker for smaller masses and later redshifts.
Hence it is not surprising that both the MW96 and SB02 models do well
at reproducing these trends, although again, only the SB02 model is
applicable if either $\nu_1$ or $\nu_2 \leq 1.$  (The P98 model is
omitted here as it was constructed only at a single redshift.)

At smaller separations, the two-redshift case differs significantly
from the behavior seen in Figure \ref{fig:xiall}.  As it is impossible
for an object to have exactly the same mass at two significantly
different redshifts, each of the numerical values approaches $\xi_L =
- 1 $ at small separations.  Thus, just as the  MW96 model
underestimates the (formally infinite) small-distance clustering of
two halos of the same mass and redshifts, it vastly overestimates the
small-distance clustering of two halos with the same mass and
different $z$ values.  On the other hand, the SB02 model is
constructed so as to exclude the formation of a halo of the same mass
at the same position at multiple redshifts, and thus the resulting
curves turn over at small separations, following the simulation.  In
fact, good agreement is seen at all separations, masses and redshifts,
with the possible exception of a slight underestimate in the cross
correlation between low mass $z_1=1$ halos and those at $z_2=5$, a
comparison which includes both the most nonlinear (lowest $\nu$ value)
peaks and the largest offset between $z_1$ and $z_2$.

\subsection{Two Redshifts, Two Masses}

\begin{figure}
\centerline{ \psfig{file=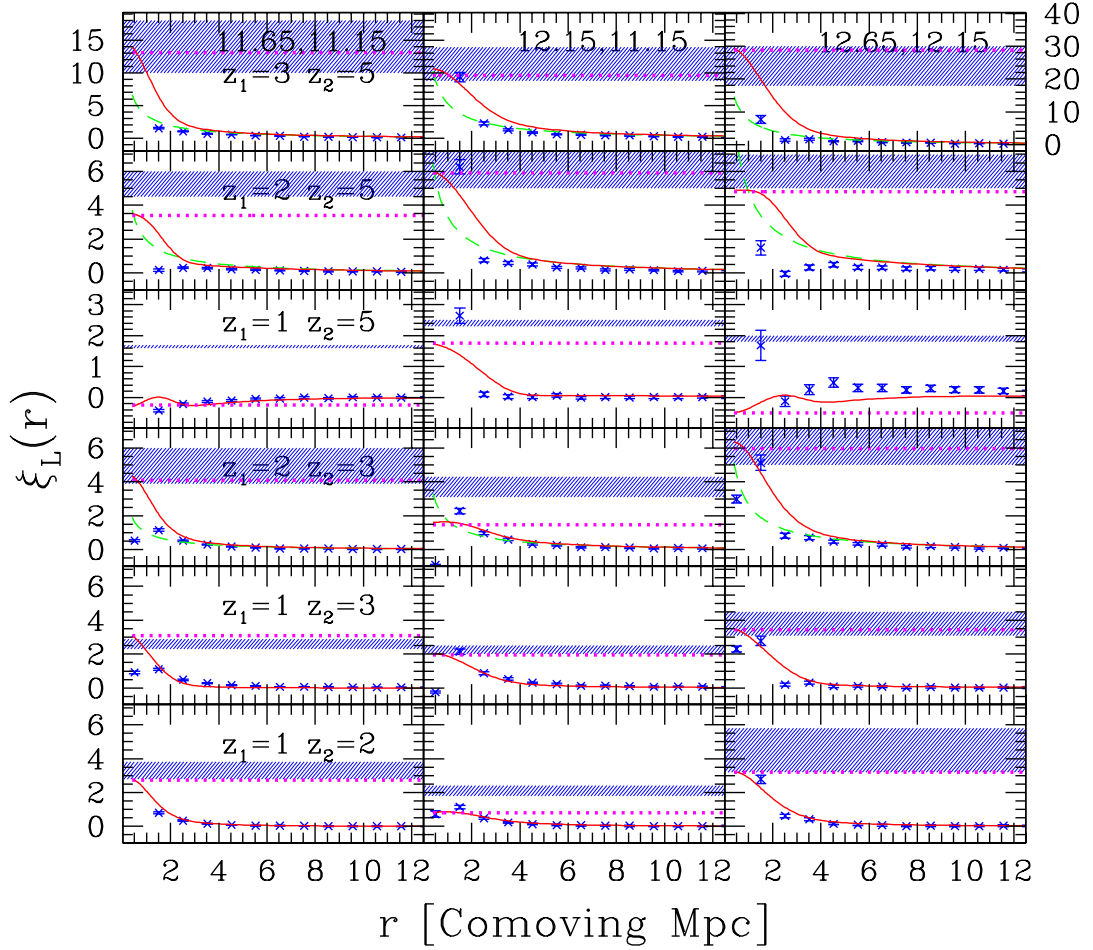,width=6.2in}}
\caption{Lagrangian correlation function of objects with different
masses and redshifts, with the smaller mass at the higher redshift.
In this figure, the points are the simulation results, the dashed and
solid curves are the MW96 and SB02 models, and the horizontal dotted
lines are the $r=0$ estimates as given by the Lacey \& Cole (1993)
progenitor distribution.  Finally, the shaded bands are the range of
progenitor number densities measured from the simulations, as
described in \S3.4. Each row is labeled by its $z$ values and each
column (from left to right) corresponds to mass bins centered on
($10^{11.15} \msun$, $10^{11.65} \msun$), ($10^{11.15} \msun$,
$10^{12.15} \msun$), and ($10^{11.65} \msun$, $10^{12.15} \msun$).
The $y$-axis label on the right only applies to the upper panel at the
extreme right.}
\label{fig:xiall4}
\end{figure}

Finally, we turn our attention to the most difficult case to model
analytically, the correlation between two objects that differ both in
mass and formation redshift.  First we consider the case in which the
smaller object forms at a higher redshift.  This is essentially a
generalization of the progenitor problem, in which one is interested
in the probability that an object of mass $M_{\rm progen}$ was found
at a redshift $z_{\rm progen}$ at a position at which an object with a
mass $M_{\rm final}$ is known to exist at a redshift $z_{\rm final}$.
This distribution was first constructed analytically in the context of
the Bond \etal (1991) excursion set formalism by Lacey \& Cole (1993),
in what has now become a standard approach.  As the SB02 model was
also developed within this context, it matches this model exactly at
$r=0$, as is clear from Figure \ref{fig:xiall4}.

Also in this figure, we see that the SB02 expression does well at
larger separations for all combinations of masses and redshifts. At
small separations, however, a significant discrepancy is seen.  As in
Figure \ref{fig:xiall}, this is primarily due to pushing our
comparison into separations in which $r \leq R_L(M).$ At these short
distances, our numerical and analytical approaches are essentially
calculating different quantities.  From a numerical perspective, our
choice of bins with a 1 Mpc width means that changes in the position
of the Lagrangian center of mass of the smaller object {\em within}
the larger object are able to move power between the leftmost bins.
On the other hand, the analytical estimates are constructed to
reproduce the total number of $M_2$ halos merging into $M_1$,
regardless of where they lie within the final object.  Thus a fair
comparison between our analytical and numerical results can only be
carried out by choosing an inner bin-width that is representative of
the size of final halo.

\begin{figure}
\centerline{ \psfig{file=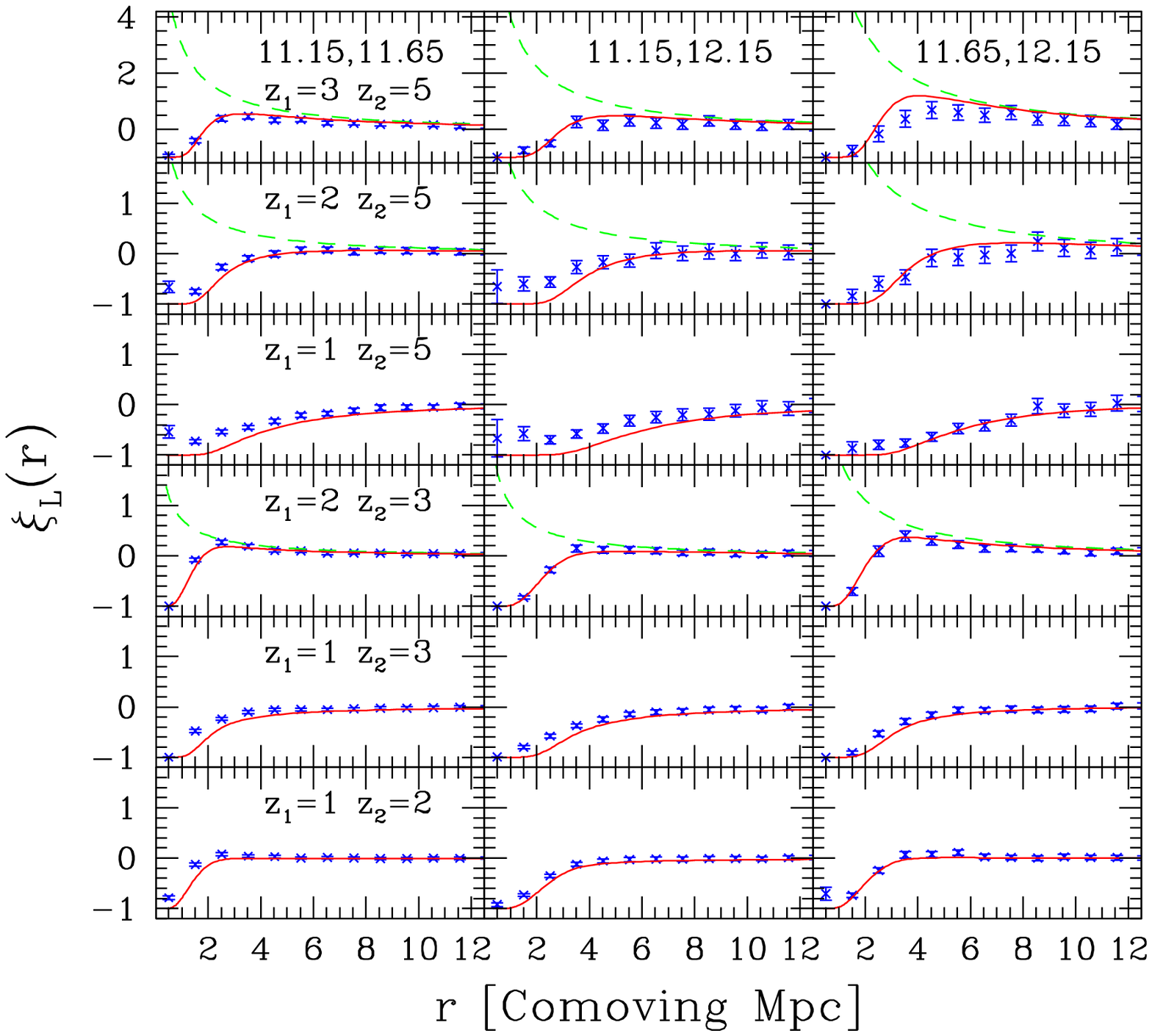,width=6.2in}}
\caption{Lagrangian correlation function of objects with different
masses and redshifts, with the larger mass at the higher redshift.
Rows and columns are as in Figure  \protect\ref{fig:xiall4}.}
\label{fig:xiall5}
\end{figure}

As there is some flexibility in this definition, we recalculated the
numerical correlation function at zero separation over a range of
bin-widths from $R_L(M_1)$ to $R_L(M_1) + R_L(M_2)/2,$ where $M_1$ is
the mass of the larger halo.  The resulting range of $\xi_L(0)$ values
are represented by the shaded regions in Figure \ref{fig:xiall4}.  In
general the larger the bin-width, the weaker the overall correlation
function.  Thus, for example, the $z_1=3$, $z_2=5$, $M_1 = 10^{11.65}
\msun$, $M_2 = 10^{11.15} \msun$ model attains $\xi_L(0)$ values of
18.7 to 10.1 over the range of bin-widths from $1.5$ to $2.0$ comoving
Mpc respectively.  Comparing these with the Lacey \& Cole (1993)
estimates we find that our numerical results are generally in
agreement if the difference between redshifts and masses is small.
For larger differences in mass and redshift, however, significant
discrepancies exist between these approaches, and there is no obvious
trend with mass and redshift that might suggest an underlying
relationship in the results.

Similar inaccuracies have been observed elsewhere in the literature.
For example in Somerville \etal (2000) (Fig 2), differences of $\sim
2$ between the analytical and numerical results are seen in cases in
which $D(z_1)/D(z_2) \sim 2$ or $M_1/M_2 \gtrsim 3,$ while order of
magnitude discrepancies can arise if $D(z_1)/D(z_2) \geq 3.$ Thus the
trends seen in this figure represent the limitations of pushing the
standard progenitor expression outside of the $M_1/M_2$ and
$D(z_1)/D(z_2)$ ranges in which it was originally tested (Lacey \&
Cole 1994). Clearly more sophisticated models for progenitor
distributions are necessary to reproduce this behavior (\eg Manrique
et al. 1998; Chiueh \& Lee 2001; Sheth \& Tormen 2002; Sheth 2003).

In Figure \ref{fig:xiall5} we consider the inverse progenitor problem,
assigning an earlier formation redshift to the object with the greater
mass.  As gravitation is only able to increase the mass of any
structure with time, this means that no such pair can exist at $r=0$
and thus $\xi_L$ must approach $-1$ at small separations.  In this
case, the subtle $r=0$ behavior seen in Figure \ref{fig:xiall4} is
replaced by simple exclusion, and the simulated results are well
reproduced by the SB02 model for all choices of parameters.  As this
turn-over requires $\xi_L(r)/\xi_{\rm DM} (r)$ to vary with radius,
the simple MW96 estimates are unable to reproduce this behavior, and
thus are widely discrepant for small $r$ values.

\section{Conclusion}

Despite the widespread use of single-point ``merger-tree'' approaches
in analytical studies of galaxy formation, nonlocal processes abound
in cosmology. 
Starburst galaxies have drove winds strong enough to strip the
gas out of neighboring density perturbations.  The resulting
inhomogeneous enrichment of the IGM profoundly affected the cooling
properties of later-forming galaxies,  perhaps altering their stellar
initial mass function. The ultraviolet radiation from  the earliest
objects strongly inhibited the formation of similar such sources.
Indeed, the moment the universe emerged from the cosmological ``dark
ages'' it was plunged into an era of strong, nonlocal feedback that
continues to this day.

In this study, we have used high-resolution N-body simulations to
examine the Lagrangian dark-matter halo correlation function, a
quantity that is uniquely suited to studying nonlocal processes
because: (1) it
is in the natural coordinate systems for PS-type analytical
calculations; (2) it allows for a more accurate treatment of the
propagation of cosmological disturbances, which are largely dependent
on the total column depth of material  separating two perturbations.
Comparing our results with a number of analytic expressions, we find
that the bivariate mass function  derived in SB02 can be used to
accurately reproduce the  Lagrangian correlation function of halos
with different  masses and formation redshifts.  This is a significant
step forward,  which confirms that non-local processes can be modeled
within the framework of the bivariate mass function.

In the case of single mass, single redshift halos, the SB02 model and
the P98 model both reproduce the correlation function behavior
well. As  expected we confirm the failure of the MW96 model for low
$\nu$ and small  radii ($r< 4$ Mpc), but the $b_E=b_L+1$ mapping
proposed by Jing works well  for all masses and radii, unless
$\nu$\lsim$1.5$.  At this point the mapping  breaks down, and a more
sophisticated technique is needed to deal with  the short-range
physics that comes into play.

For objects of different masses at the same redshift, the P98 model
and the  SB02 model have close agreement to a limiting radius that  is
defined by the turn-over in the correlation function due to the
anticorrelation required at zero separation. The P98 model does not
correctly reproduce this turn-over, due to the omission of the
appropriate barrier treatment for the formation of structure within
peaks, while the SB02 model exhibits the correct short-range  behavior
for all our studied cases of different masses and redshifts.

The different mass, same redshift Eulerian correlation functions for the
SB02 and P98 models exhibit similar performance issues to the same mass,
same redshift comparison. The problems are particularly acute for the
SB02 model where the barrier treatment introduces a strong kink in the
Eulerian correlation function. In these cases, remarkably, the Eulerian
MW96 model works best of all three models.

The two redshift, equal mass Lagrangian case also shows the SB02 model
to perform well. The growth of halo mass with redshift forces
anticorrelation of equal mass halos at different redshifts and small
radii.  The correct turnover in the correlation function for small
radii is  observed, and the position of the knee is accurately
predicted for all  redshift cases.

Unsurprisingly, our most interesting results come from the most
difficult  case to study, namely the correlation of halos with
different masses and  formation redshifts. In the case in which the
larger mass is at the higher redshift, the SB02 model reproduces the
expected anticorrelation extremely  accurately. In the more relevant
case of the lower mass at higher redshift  the Lacey \& Cole (1993)
progenitor distribution is always reproduced exactly at small radii,
and this distribution compares well with the simulations,  provided
that the differences in redshift and mass between the halo are not too
large ($M_2/M_1<10$ and $z_1-z_2<3$).  For larger differences in mass
and redshift, however, both the Lacey \& Cole (1993) and SB02 models
exhibit significant discrepancies from the numerical values measured
here and reported elsewhere.  At larger radii, however, the SB02 model
reproduces the expected  behavior of the correlation function in all
cases.  It is intriguing that the {\em single-point} case  remains the
single most important outstanding issue in our analytic modeling of
Lagrangian bias.

\acknowledgements

ES would like to express his sincere thanks for the hospitality shown
to him by J. Richard Bond and the Canadian Institute for Theoretical
Astrophysics (CITA), where much of this research was carried out.
RJT\ acknowledges funding from the Canadian Computational Cosmology
Consortium and use of the CITA and SHARCNET computing facilities.  We
are grateful to Rennan Barkana and an anonymous referee for helpful
comments.  This work was supported by the National Science Foundation
under grant PHY99-07949.

\fontsize{9}{9pt}\selectfont

\end{document}